\documentclass{aa}  

\usepackage{graphicx}
\usepackage{txfonts}
\usepackage{xcolor}
\newcommand{\nndp}{$\rm N_2D^+$\xspace}

\newcommand{\nnhp}{$\rm N_2H^+$\xspace}
\newcommand{\kms}{$\rm km \, s^{-1}$\xspace}

\newcommand{\cdo}{$\rm C^{18}O$\xspace}
\newcommand{\am}{$\rm NH_3$\xspace}

\begin{document}

   \title{Hunting pre-stellar cores with APEX: Corona Australis 151, the densest pre-stellar core or the youngest protostar?}

\author{E. Redaelli\inst{1,2} \and S. Spezzano\inst{2}   \and P. Caselli\inst{2} \and J. Harju \inst{3,2} \and D. Arzoumanian \inst{4} \and O. Sipil\"a\inst{2} \and A. Belloche\inst{5} \and F. Wyrowski\inst{5} \and J. E. Pineda \inst{2}}  

\institute{European Southern Observatory, Karl-Schwarzschild-Strasse 2, 85748 Garching, Germany \and Max-Planck-Institut f\"ur Extraterrestrische Physik, Giessenbachstrasse 1, 85748 Garching, Germany \and Department of Physics, P.O. Box 64, FI-00014, University of Helsinki, Finland  \and National Astronomical Observatory of Japan, Osawa 2-21-1, Mitaka, Tokyo 181-8588, Japan \and
        Max-Planck-Institut f\"ur Radioastronomie, Auf dem H\"ugel, 69, 53121 Bonn, Germany
}

   \date{XXX}

 
  \abstract
   {Pre-stellar cores are the birthplaces of Sun-like stars and represent the initial conditions for the assembly of protoplanetary systems. Due to their short lifespans, they are rare. In recent efforts to increase the number of such sources identified in the Solar neighbourhood, we have selected a sample of 40 starless cores from the publicly available core catalogs of the \textit{Herschel} Gould Belt survey. In this work, we focus on one of the sources that stands out for its high central density: Corona Australis 151.}
   {We use molecular lines that trace dense gas ($n \gtrsim 10^6\, \rm cm^{-3}$) to confirm the exceptionally high density of this object, to study its physical structure, and to understand its evolutionary stage.}
   {We detected the \nnhp $3-2$ and $5-4$ transitions and the \nndp $3-2$, $4-3$, and $6-5$ lines with the Atacama Pathfinder EXperiment (APEX) telescope. We use the \textit{Herschel} continuum data to infer a spherically symmetric model of the core's density and temperature. This is used as input to perform non-local-thermodynamic-equilibrium radiative transfer to fit the observed five lines.}
   {Our analysis confirms that this core is characterized by very high densities (a few $\times 10^7 \, \rm cm^{-3}$ at the centre) and cold temperatures ($8-12\, \rm K$). We infer a high deuteration level of $\rm N_2D^+ / N_2H^+ = 0.50$, indicative of an advanced evolutionary stage. In the large bandwidth covered by the APEX data, we detect several other deuterated species, including $\rm CHD_2OH$, $\rm D_2CO$, and $\rm ND_3$. We also detect multiple sulphurated species that present broader lines with signs of high-velocity wings.}
       {{High-angular resolution observations will be necessary to unveil the evolutionary stage of Cra 151. The detection of a compact emission at 70 $\mu$m does not
exclude that the source is a First Hydrostatic Core (FHSC) or in a very early stage of the protostellar 
phase. The observation of high-velocity wings and the fact that the linewidths of \nnhp and \nndp become larger with increasing frequency can be interpreted either as an indication of supersonic infall motions developing in the central parts of a very evolved pre-stellar core or as the signature of outflows from a very low luminosity object (VeLLO).}}

  \titlerunning{Hunting pre-stellar cores with APEX: CrA 151}
\authorrunning{E. Redaelli et al.}
  \maketitle

%

\section{Introduction}

Pre-stellar cores are the preferred location for star formation. These dense and cold fragments of molecular clouds represent the initial conditions for the assembly of stellar systems, and their study is hence crucial to understand the subsequent planetary formation. These objects are however rare, due to their short lifetimes before the gravitational collapse. L1544, embedded in the Taurus molecular cloud, represents a prototypical example of a pre-stellar core. Recent observations with the Atacama Large Millimeter and sub-millimeter Array \citep[ALMA][]{Caselli19,Caselli22} showed the existence at its centre of a high-density "kernel" (size: $\lesssim 2000\,$au, temperature $6\, \rm K$, density $n> 10^7 \, \rm cm^{-3}$) which is in agreement with the predictions of theoretical works focussing on the dynamical evolution of contracting magnetized clouds \citep[e.g.][]{Galli93}. \par
In recent observational efforts, we have sought to expand the catalogue of known pre-stellar cores, to increase our understanding of their physical and chemical properties statistically. We have identified a sample starting from the {\em Herschel} Gould Belt Survey Archive (HGBS\footnote{All data available at \url{http://gouldbelt-herschel.cea.fr/archives}.}; \citealt{Andre10}), and we have targeted it with single-dish observations of high frequency ($230-460\,$GHz) lines of \nnhp and \nndp. The whole sample, which is constituted by the densest starless cores identified in the HGBS within 200$\,$pc from the Solar System, is presented and described in Caselli et al. (in prep.). The high-$J$ transitions of \nnhp and \nndp can be used to confirm the high central density, due to their high critical densities ($n_\mathrm{crit} \gtrsim 10^6 \, \rm cm^{-3}$). Furthermore, the deuteration level of diazenylium $R_\mathrm{D} = N(\text{\nndp})/N(\text{\nnhp})$ is a well-known evolutionary indicator for the pre-stellar phase \citep[cf.][]{Crapsi05}. These species are formed from molecular nitrogen, which chemically is a late-type molecule (cf. e.g. \citealt{Hily-Blant10}). Furthermore, the deuteration process is efficient at advanced evolutionary stages, as it is enhanced by low temperatures and high CO depletion factors (\citealt{Dalgarno84}; see also the review from \citealt{Ceccarelli14}). Using the low-$J$ transitions of \nnhp and \nndp, \cite{Crapsi05} found that a high deuteration level ($R_\mathrm{D}>0.1$) is a sign of a centrally concentrated core with a high degree of CO depletion.
\par
We identified from the sample the objects where the high rotational transitions (in particular \nndp $6-5$ and \nnhp $5-4$) have been successfully detected, to study them in detail. The present work focuses on the core identified as Corona Australis 151 (hereafter, CrA 151) in the HGBS core catalogue of \cite{bresnahan18}, found in the CrA-E region. A companion work analyses the core Ophiuchus 464, also known as IRAS16293E \citep{spezzano24}. 
\par
The Corona Australis molecular cloud is a well-studied low-mass star-forming region in the Southern Hemisphere. Its morphology has been studied in detail using \textit{Herschel} data as part of the HGBS survey \citep{bresnahan18}. Its shape resembles a cometary head in the Western part, and two filamentary structures stretching towards the East. Star formation is particularly active in the Western end, which hosts the so-called Coronet cluster. This region is known to host several Young Stellar Objects (YSOs) at distinct evolutionary stages (cf. \citealt{Chini03, Sicilia-Aguilar13}). Star formation appears less active in the rest of the cloud.
The most recent distance estimate for Corona Australis is $d=150\, \rm pc$, based on GAIA data \citep{Galli20}. We adopt this value throughout this work.
\par 
The core CrA 151 sits in the northern filament of \citet[][region CrA-E]{bresnahan18}, at about $ 5 \, \rm pc$ East of the Coronet cluster. It corresponds to Cloud 42 of \cite{Sandqvist76}. The region contains $\approx 20 \, \rm M_\odot$ of total mass and one robust pre-stellar core candidate (CrA 151 itself). No young star, disc, or YSO is known within the core boundaries { according to the census reported in \cite{Esplin22}, which is based on proper motions measured from multi-epoch IR imaging from the Spitzer Space Telescope and Magellan Observatory and positions in the color-magnitude diagram measured with Gaia, 2MASS, VISTA VHS, WISE, Spitzer, and Magellan. However, \cite{bresnahan18} detected a point source at 70$\, \mu$m towards Cra 151 with $Herschel$/PACS. While emission at 70$\,\mu$m is not conclusive proof of the presence of a central heating source, and the detection is at the 4$\sigma$ level, it might indicate the presence of a Very Low Luminosity Object (VeLLO; \citealt{difrancesco07, tomida10}), which could be a very young low-mass protostar or a First Hydrostatic Core (FHSC, \citealt{larson69})}.
\cite{Hardegree-Ullman13} studied the region using \cdo and \nnhp $1-0$ observations from the Swedish ESO Submillimeter Telescope (SEST). Those authors modelled the density profile of the core using a Plummer-like solution using the available continuum data, finding a central density higher than $10^6 \rm \, cm^{-3}$. They ran chemical modelling to investigate the core's chemistry and found evidence of strong CO depletion. \nnhp, instead, appears to have a rather constant abundance of a few $\times 10^{-10}$. However, their line observations had limited angular resolution ($\sim 50''$) and, targeting the low-$J$ transitions of two abundant species, they were more sensitive to more extended regions within the core than our observations (see below). \par
This work is organised as follows. Section~\ref{Observations} describes the observational data. The obtained spectra and subsequent analysis are presented in Sect.~\ref{sec:analysis}, where first we develop the physical model of the source (Sect.~\ref{sec:physmod}) and then we present the radiative transfer modelling of the spectroscopic lines (Sect.~\ref{sec:LOC}). Section~\ref{sec:otherlines} describes other species detected in the frequency coverage of the observations. We discuss the results of the analysis in Sect.~\ref{sec:discussion}. Section~\ref{sec:conclusions} contains a summary and final remarks of this work. 
\begin{table*}[!t]
    \renewcommand{\arraystretch}{1.4}
\centering
\caption{Parameters of the targeted lines.}
\label{tab:lines}
\begin{tabular}{c|ccccccccc|cc}
       \hline \hline
Transition & Frequency\tablefootmark{a} & $A_\mathrm{ul}$\tablefootmark{a} & $g_\mathrm{u}$\tablefootmark{a} & $E_\mathrm{u}$\tablefootmark{a} & $n_{\rm crit}$\tablefootmark{b} & $\theta_\mathrm{MB}$ &  $\eta_\mathrm{MB}$&$\Delta V_\mathrm{ch}$ & $rms$ & $V_\mathrm{lsr}$\tablefootmark{c} & $FWHM$\tablefootmark{c} \\
 		& MHz	& $\times 10^{-3} \; \rm s^{-1}$	&		& K			& $\rm 10^6 \, cm^{-3}$	&	$''$			&			& \kms & mK  & \kms & \kms \\

\hline
\nnhp $3-2$ &  279511.7491  & $1.26$             &   63      &  26.8 &   1.4   	& 22.3&	0.76& 0.07     &   52        &   5.612(5) & 0.397(13)\\
\nnhp $5-4$ &  465824.7770 & $6.18$              &    99     &  67.1 &   5.5   	&13.3	&	0.60&0.04  & 33   &  5.661(6)   & 0.747(11)\\
\nndp $3-2$ & 231321.8283  & $7.14 \times 10^{-1} $ &  63    &  22.2&   0.74    	&26.0	&	0.80&0.08  & 9     &   5.6543(2)    & 0.3889(12)\\
\nndp $4-3$ & 308422.2672  & $1.75$               &   81     &   37.0&	 1.5      &20.2	&	0.74	& 0.06   & 37     &   5.663(6)  & 0.43(3)\\
\nndp $6-5$ &  462603.8520 & $6.15$              &    117     &  77.7&   4.4    	&13.5	& 0.60		& 0.04 & 29     &  5.61(3)     & 0.57(7)
\\
 \hline
\end{tabular}
\tablefoot{\tablefoottext{a} {The spectroscopic information, i.e. the line frequency, the Einstein coefficient for spontaneous emission $A_\mathrm{ul}$, the upper-level degeneracy $g_\mathrm{u}$ and energy  $E_\mathrm{u}$ are taken from the CDMS catalogue \citep{Muller05}. They refer to the unsplitted rotational transitions.} \\
\tablefoottext{b} {The line critical density is computed with Eq. 4 of \cite{Shirley15}, using the collisional rates of the unsplitted lines.} \\
\tablefoottext{c}{ Kinematic parameters obtained with the \textsc{hfs} fit. The numbers in paranthesis are the uncertainties on the last digit.}}
\end{table*}
\section{Observations\label{Observations}}
The data of the diazenylium lines analysed in this work were collected with the Atacama Pathfinder EXperiment (APEX, \citealt{Gusten06}) telescope as part of the large core sample described in Caselli et al. (in prep.). We refer to that paper for an exhaustive description of the observational details. Here, we offer a summary for the particular case of CrA 151. \par
The data consist of single-pointing observations towards the core's centre, at coordinates $\rm RA(J2000) = 19^{\rm h}10^{\rm m}20^{\rm s}.17$ and $\rm Dec(J2000)= -37^{\circ}08^{\rm m}27^{\rm s}.0$.
Two tunings of the receiver SEPIA345 were used to cover the \nnhp $3-2$ and \nndp $4-3$ transitions at 279.5$\,$GHz and 308.4$\,$GHz, respectively. These observations were performed in July and October 2022 (Proposal ID: O-0110.F-9310A-2022). The \nndp $3-2$ line was covered with the nFLASH230 receiver, whilst the \nndp $6-5$ and \nnhp $5-4$ transitions were covered simultaneously with one tuning of nFLASH460 (we exploited the dual-channel capability of the nFLASH frontend, \citealt{Klein14}). These observations were collected under project M-0110.F-9501C-2022 in September and November 2022. \par
In all cases, the observations were performed in ON-OFF position switching, with OFF coordinates $
\Delta \rm RA,\Delta Dec = (-300'', 0)$ with respect to the core central coordinates. We used the FFTS backend, which allows a spectral resolution of 64$\,$kHz. This translates into velocity resolution from $\Delta V_\mathrm{ch}\sim 0.08\,$\kms at 230$\,$GHz to $\sim 0.04\,$\kms at 460$\,$GHz. \par
We reduced the data using the \textsc{class} package of the \textsc{gildas} software\footnote{Available at \url{http://www.iram.fr/IRAMFR/GILDAS/}.}. In particular, we subtracted baselines (usually a first-order polynomial) in every  scan, and then averaged them. The intensity scale was converted into main beam temperature $T_\mathrm{MB}$ assuming the forward efficiency $F_\mathrm{eff} = 0.95$ and computing the main-beam efficiency at each frequency according to $\eta_\mathrm{MB} = 1.00797-0.000857*\nu(\rm GHz)$. We obtained this relation fitting the available data tabulated at \url{https://www.apex-telescope.org/telescope/efficiency/index.php}. Relevant information on the targeted transitions and their observational parameters is given in Table~\ref{tab:lines}, including the angular resolution ($\theta_{\rm MB}$) of the observations, which spans the range $13-26''$. At the distance of CrA 151, the linear resolution is $2000-4000\,$au.

\section{Results and Analysis \label{sec:analysis}}
This Section contains a description of the observed spectra of \nnhp and \nndp and the analysis we performed to model them. It concludes with a discussion of the other lines that were covered by the APEX data.

\subsection{Observed spectra\label{sec:results}}
\begin{figure}[!h]
    \centering
    \includegraphics[width=\linewidth]{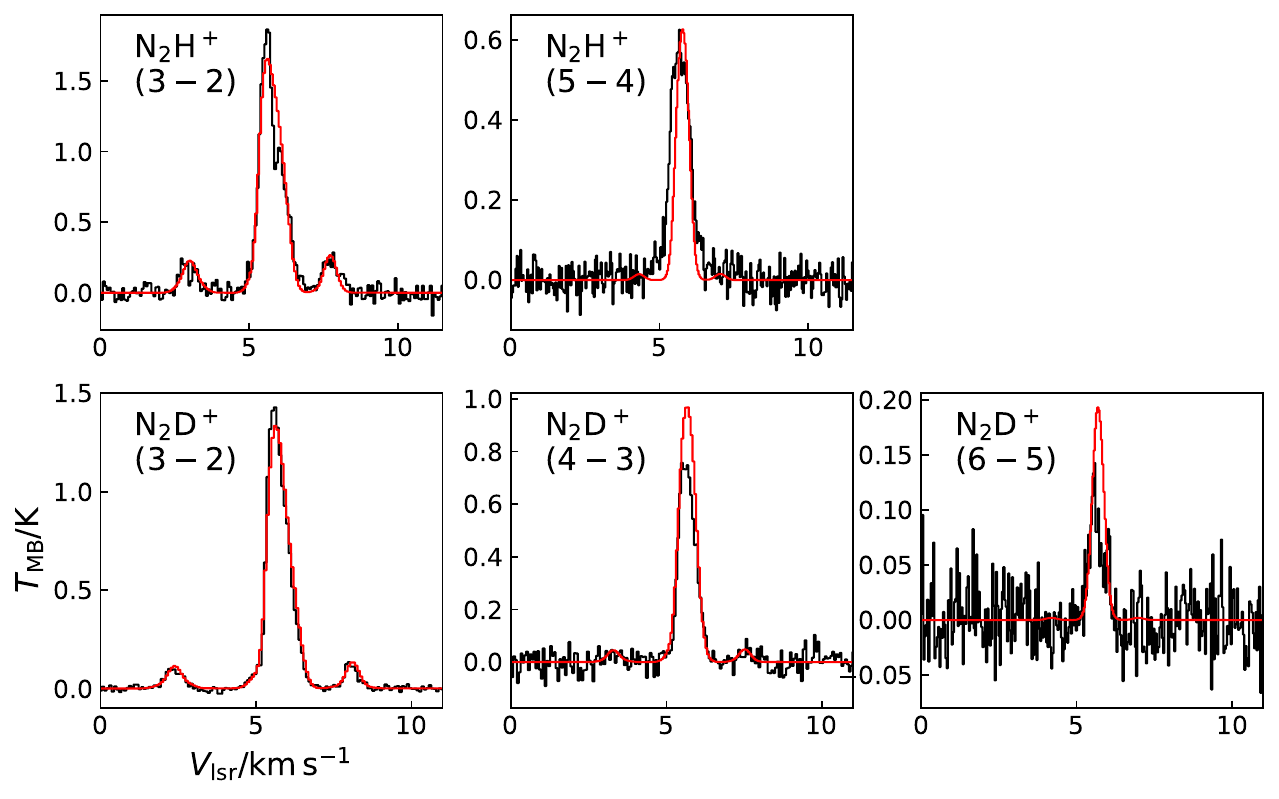}
    \caption{Black histograms show the observations of the \nnhp and \nndp transitions towards CrA 151. The red histograms show the synthetic spectra obtained with the MCMC approach applied to the non-LTE radiative transfer analysis (see Sect.~\ref{sec:LOC}).}
    \label{fig:n2hp_n2dp}
\end{figure}

The plots of the observed \nnhp and \nndp transitions are shown in Fig.~\ref{fig:n2hp_n2dp}. All the lines are robustly detected based on the sensitivities reached ($rms$ reported in Table~\ref{tab:lines}). The peak signal-to-noise ratios (S/N) span from $\rm S/N = 5 $ (\nndp $6-5$) to $\rm S/N = 160$ (\nndp $3-2$). We compute the critical densities of these transitions, based on our own calculation from the collisional and Einstein rates of the pure rotational transitions (see Table~\ref{tab:lines}). The values range from $7.4 \times 10^5$ to $5.5 \times 10^6\, \rm cm^{-3}$. Their detection with a single-dish facility at a resolution of $4000 \, \rm au$ (at the lowest frequency) is indicative that high densities characterise a significant portion of the core. This point will be discussed further in Sect.~\ref{sec:physmod}.
\par

We have fit the data assuming a constant excitation temperature for all the hyperfine components within each rotational transition, using the \textsc{hfs} routine implemented in \textsc{class} to obtain the kinematic parameters of the lines, namely the local-standard-of-rest velocity ($V_\mathrm{lsr}$) and the line full-width-half-maximum ($FWHM$). The latter is the instrinsic linewidth, not biased by  opacity broadening. The obtained best-fit parameters are listed in the last two columns of Table~\ref{tab:lines}, and the corresponding plots are shown in Appendix~\ref{app:CLASSFit}. The $V_\mathrm{lsr}$ values obtained for \nnhp $5-4$ and all the \nndp transitions agree within $3\sigma$, around a weighted average of $\langle V_\mathrm{lsr} \rangle = (5.6543 \pm 0.0002)\,$\kms. The value for \nnhp $3-2$ is lower: $(5.612 \pm 0.005) \, $\kms. The difference of $\sim 0.04 \,$\kms is significant above the $5 \sigma$ level. However, we highlight that this difference is smaller than the channel width at $1\,$mm. Furthermore, the \nnhp $3-2$ transition shows a clear profile feature (the double peak seen in the central hyperfine group) that is not due to the hyperfine structure and that cannot be reproduced by the \textsc{hfs} fitting routine (cf. Fig.~\ref{fig:CLASSFit}). This likely arises from gas motions within the core, possibly due to gravitational contraction, which results in double-peaked profiles with blue asymmetry in optically thick lines \citep{Evans99}. The total optical depth of this transition, estimated by \textsc{class}, is $\tau_\mathrm{tot} = 4.2\pm 0.3$. The presence of this feature affects the determination of $V_\mathrm{lsr}$ for this line. \par
Concerning the linewidths, the transitions at $\sim 1 \,$mm present narrow lines ($FWHM \approx 0.4 \,$\kms). The \nndp $6-5$ line appears broader, even though its $FWHM$ has large uncertainties due to the low $\rm S/N$ of the spectrum. On the other hand, the \nnhp $5-4$ line is significantly broader ($FWHM = 0.747 \pm 0.011 \,$\kms).

\begin{figure*}[!t]
   \centering
   \includegraphics[width=\textwidth]{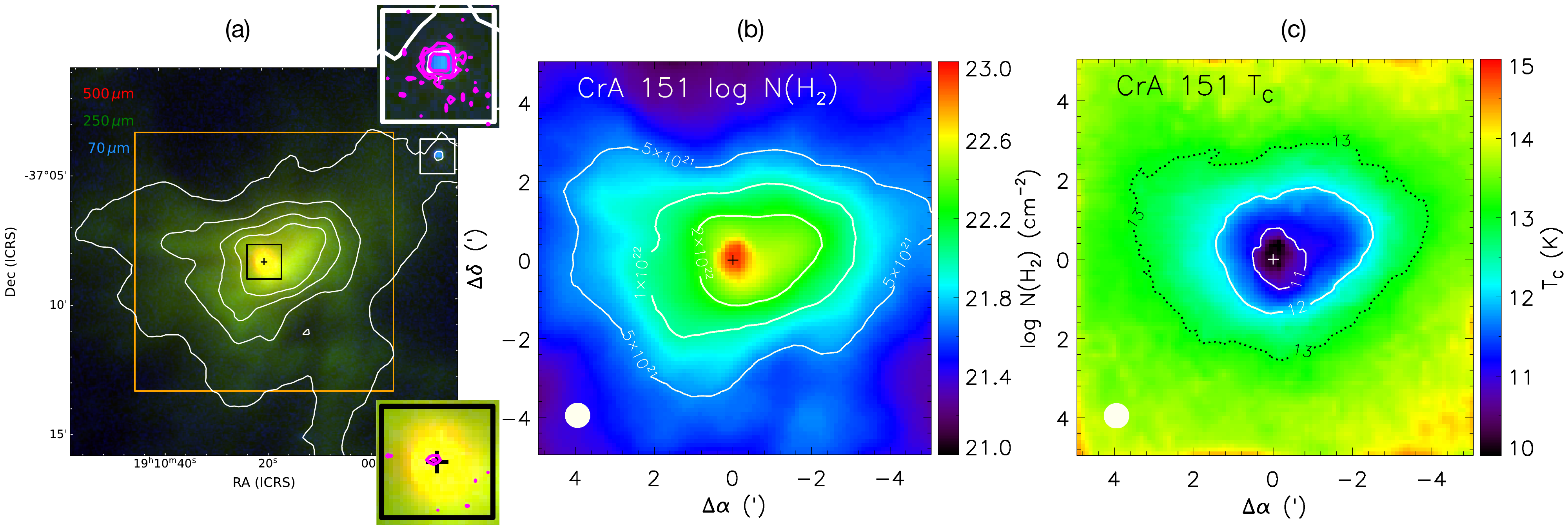}
   \caption{Panel (a): three-colour RGB image of CrA 151 obtained using the \textit{Herschel} data at $500 \, \rm \mu m$ (red), $250 \, \rm \mu m$ (green), and $70 \, \rm \mu m$ (blue). The contours show the distribution of the $350 \, \rm \mu m$ SPIRE flux. {The two square subpanels show zoom-ins of the core’s centre (bottom right subpanel) and of the area
around the YSO in the north-west of the core (top right subpanel), with the 70$\, \mu$m \textit{Herschel} emission in magenta contours (levels: $[25,50,150]\rm\, MJy/sr$; $1\sigma=15 \, \rm MJy/sr$).} The orange rectangle shows the region presented in the next two panels. Panel (b) and (c): zoom-in of the map of the $N\rm (H_2)$ distribution and of the dust temperature (or color temperature, $T_\mathrm{C}$) from the HGBS survey. The beam size is shown in the bottom-left corners. In all panels, the coordinates of the core's centre are shown with the `$+$' symbol. \label{fig:HerschelData}}
\end{figure*}


\subsection{Physical model of the source\label{sec:physmod}}
\begin{figure}[!h]
    \centering
    \includegraphics[width=\linewidth]{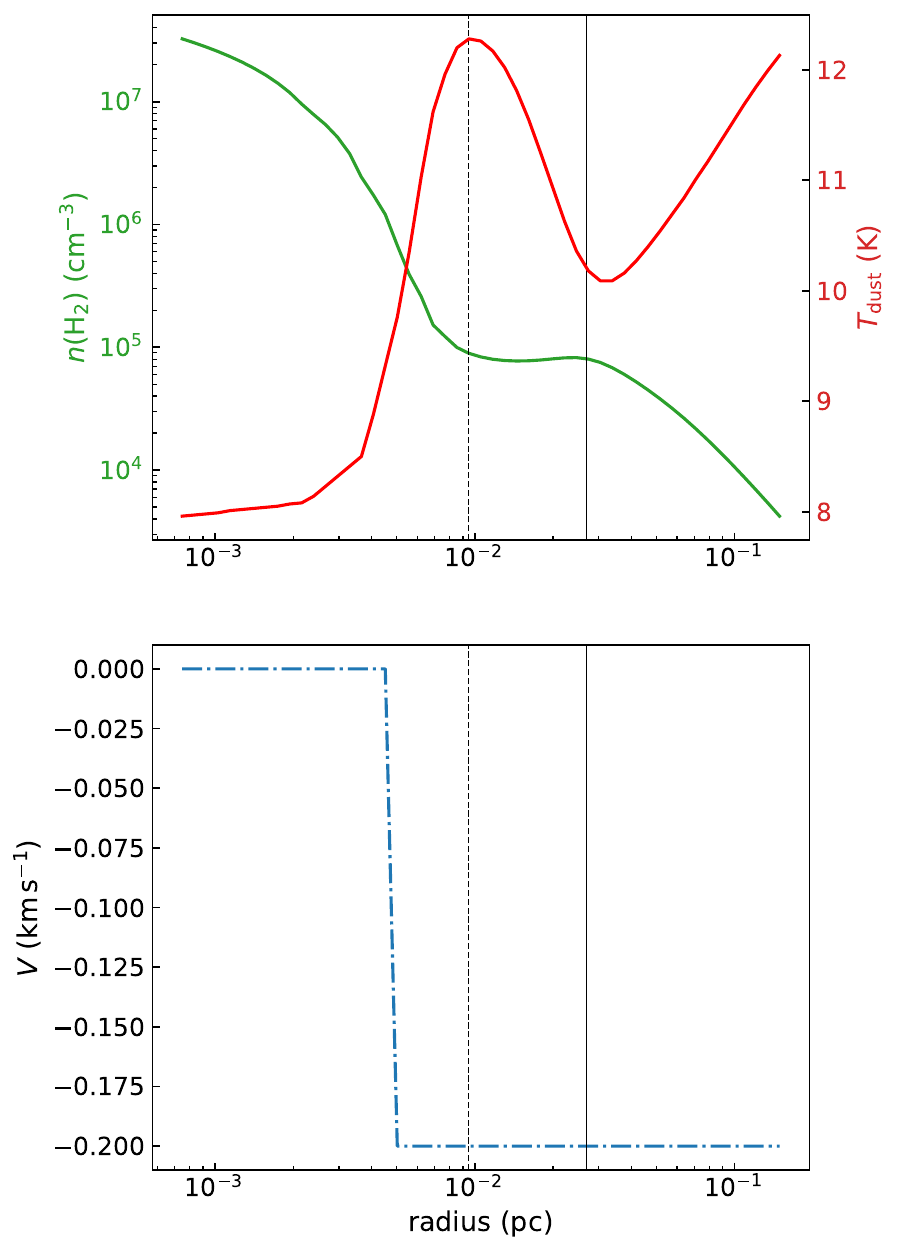}
    \caption{\textit{Top panel:} The physical model of CrA 151 obtained from the \textit{Herschel} data. The green curve (left y-axis) shows the total volume density $n\rm (H_2)$, whilst the red curve (right y-axis) shows the dust temperature. The bumps in both profiles are due to a double Plummer-like structure. \textit{Bottom panel:} The velocity profile used for the MCMC modelling. In both panels, the vertical dashed and solid lines show the minimum and maximum resolution of the observations, from $13''$ (\nnhp $5-4$) to $37''$ (\textit{Herschel} data). }
    \label{fig:cra151_model}
\end{figure}

\begin{figure}[!h]
   \centering
   \includegraphics[width=\linewidth]{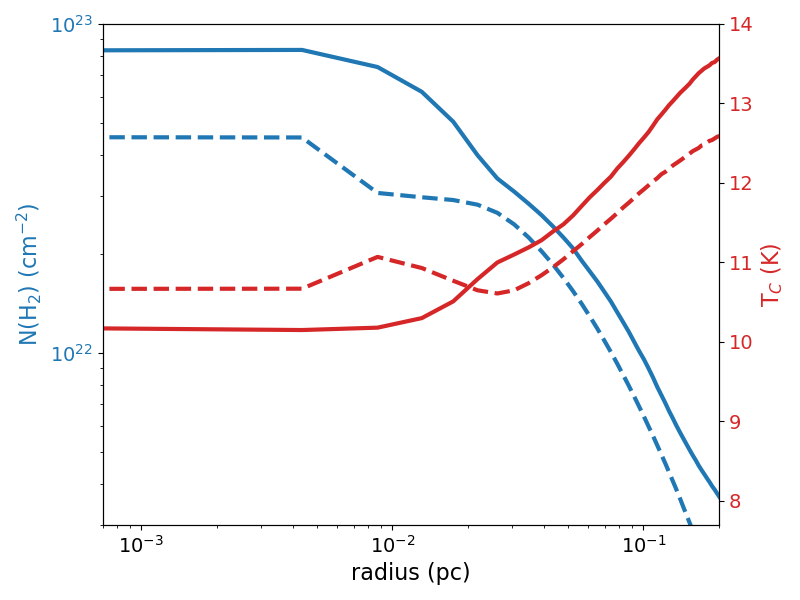}
   \caption{Radial  $N(\rm H_2)$ and $T_\mathrm{C}$ profiles of Cra 151 calculated from the $Herschel$ maps shown in Fig.~\ref{fig:HerschelData} (solid lines) and from the 1D ($n$,$T_\mathrm{dust}$) model shown in Figure~\ref{fig:cra151_model} (dashed lines).}
   \label{fig:N(H2)}
\end{figure}

To perform the non-LTE radiative transfer, we computed a spherically symmetric model of the source based on the available dust thermal emission data from \textit{Herschel}. Figure~\ref{fig:HerschelData} shows the three-color RGB image obtained from the SPIRE and PACS wavelengths in the left panel, and the derived column density and dust temperature maps in the remaining panels. The detailed procedure is described in \cite{Roy14}. Briefly, according to their Eq.\,(3), the product $\rho(r)\,B_\lambda(T_{\rm d})\,\kappa_\lambda$, where $B_\lambda$ is the Planck function at wavelength $\lambda$ and $\kappa_\lambda$ is the dust opacity at this wavelength, can be obtained from the inverse Abel transformation of the surface brightness gradient $dI_\lambda/dr$ as a function of the impact parameter. The density and temperature profiles were fitted by determining the surface brightness gradient at four wavelengths. For the dust opacity, we assumed $\kappa_{\rm 250 \mu m}  = 0.1 \, \rm cm^{2}\, g^{-1}$ and a dust opacity index of $\beta=2.0$ \citep{Hildebrand83}. We used the {\sl Herschel}/SPIRE images at $\lambda=250\,\mu$m,  $350\,\mu$m, and $500\,\mu$m, and the {\sl Herschel}/PACS image of the region at $160\,\mu$m. The maps at 160, 250, and 350$\, \mu$m were convolved to the resolution of the \textit{Herschel} beam at 500$\, \mu$m, $37''$. The surface brightness profiles were determined by fitting Plummer-type functions to the concentric circular averages. The purpose of the fitting was to provide a smooth intensity gradient $dI_\lambda/dr$. We first tested a single Plummer profile, which however cannot fit the intensity bump seen in the SPIRE 250$\, \mu$m and PACS 160$\, \mu$m profiles {(see Appendix ~\ref{app:Herschel})}. A better fit {to all wavelengths} was found using two Plummer-like profiles. The resulting volume density and dust temperature profiles are shown in Fig.~\ref{fig:cra151_model}. {Figure~\ref{fig:N(H2)} shows the radial profiles of the column density $N \rm (H_2)$ and temperature (or color temperature, $T_\mathrm{C}$) derived from the maps shown in Fig.~\ref{fig:HerschelData}b (solid lines) as well as from the 1D volume density and temperature profiles (dashed lines), as a comparison.}
\par 
The resulting temperature and density profiles, {derived using maps smoothed to $37''$}, show smooth radial gradients. The dust temperature range is $8-12\,$K, increasing outwards, indicative of cold, pre-stellar gas {at the angular resolution of the $Herschel$ observations}. The volume density values range from $10^4\, \rm cm^{-3}$ at the core's boundary ($0.15\, \rm pc$) to a few $\times 10^7\, \rm cm^{-3}$ towards the center. These high densities explain the detection of high critical-density transitions, such as \nnhp $5-4$ and \nndp $6-5$, as discussed in Sect.~\ref{sec:results}. In particular, within the central $2000\,$au (corresponding to the APEX angular resolution at 460$\,$GHz), the volume density is $n > 10^5 \, \rm cm^{-3}$.
\par
The model built so far is static. However, evolved pre-stellar cores have likely started to contract under the gravitational pull of the central overdensity. The double-peak asymmetry seen in \nnhp $3-2$ suggests this scenario. Since this line has a low critical density and a large beam size compared to the other transitions, and the \nnhp distribution is expected to be more extended than \nndp due to chemical reasons, this transition likely traces more the outer parts of the core when compared to the other lines in the sample. We have hence decided to simulate the core contraction using a step-like velocity $V$ profile, with $V=-0.2 \,$\kms for radii larger than $1000\,$au and $V=0$ elsewhere. Section~\ref{sec:discussion} contains a detailed discussion on the core's physical model and dynamical state.

\subsection{Radiative transfer analysis\label{sec:LOC}}
To model the observed spectra, we used the non-local-thermodynamic-equilibrium (non-LTE) radiative
transfer code LOC \citep{Juvela20}. We have used the hyperfine-splitted collisional rates taken from the EMAA catalogue\footnote{Available at \url{https://emaa.osug.fr/species-list}.}. The \nnhp rates were computed by \cite{Lique15}, and the \nndp ones by \cite{Lin20}. To take into account additional line broadening due to e.g. turbulence, which is not considered in the physical model built so far, we manually add a constant contribution of $\sigma_\mathrm{turb}= 0.225\,$\kms, chosen after a few tests in the range $0.075-0.3\,$\kms. Since we do not have information on the gas kinetic temperature profile, we assume $T_\mathrm{gas}=T_\mathrm{dust}$, where $T_\mathrm{dust}$ is derived from the modelling of the \textit{Herschel} data shown in Fig.~\ref{fig:cra151_model},  which is a reasonable assumption because the dust and gas coupling is expected to be efficient at high densities ($n \gtrsim 10^{4-5}\, \rm cm^{-3}$, \citealt{Goldsmith01}). \par
\begin{figure*}
    \centering
    \includegraphics[width=.8\linewidth]{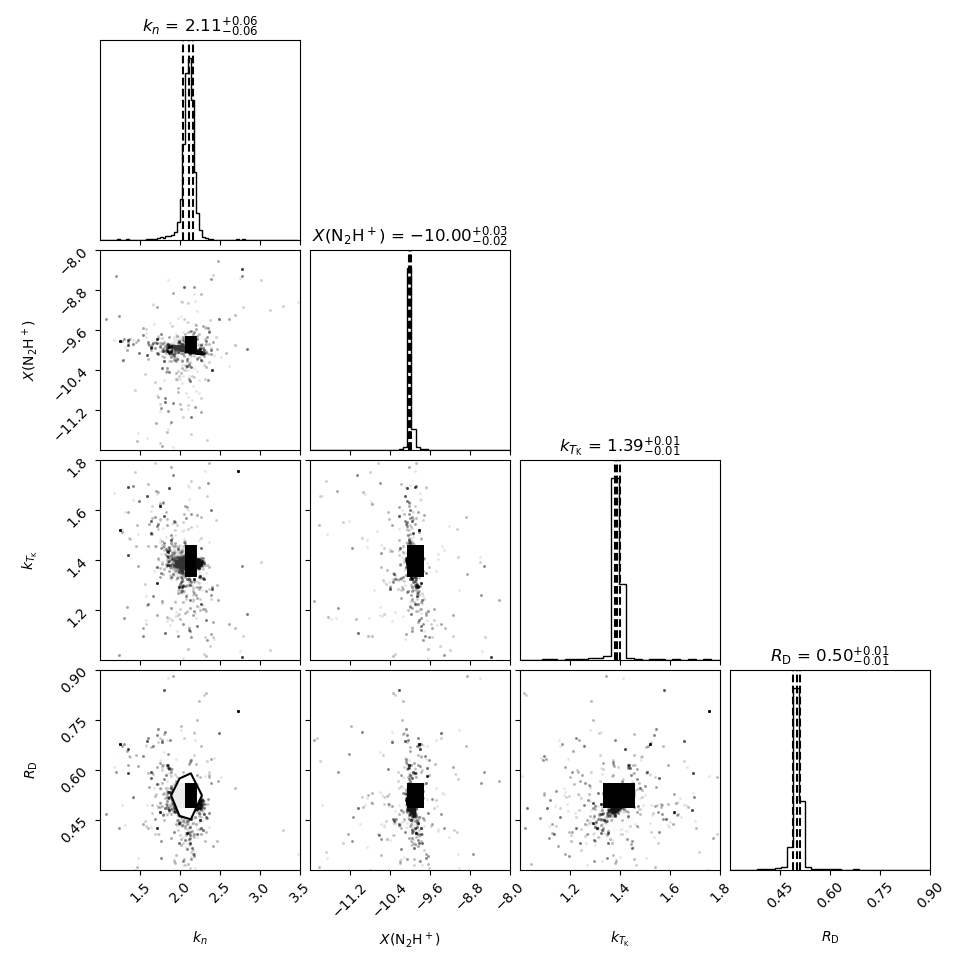}
    \caption{Corner plots of the MCMC+LOC modelling of the five \nnhp and \nndp lines towards the target. The title of each corner reports the 50th percentile (median), together with the 16th and 84th percentiles as negative and positive uncertainties, respectively. }
    \label{fig:corner}
\end{figure*}

To estimate the \nnhp and \nndp abundance, we performed a Monte Carlo Markov chain (MCMC) optimisation using the code \textsc{emcee} \citep{Foreman-Mackey23}, following the approach of \cite{Jensen24} and Ferrer-Asensio et al. (subm.). The free parameters are the constant abundance $X \rm (N_2H^+)$ of \nnhp and the constant deuteration fraction $R_\mathrm{D} = X \mathrm{(N_2D^+)}/X \rm (N_2H^+)$. In addition, to take into account the uncertainties in the physical model computation, we allowed the entire density and temperature profiles to be multiplied by factors $k_{n}$ and $k_{T \rm _K}$, respectively. We simultaneously fit the five available transitions, comparing the synthetic spectra produced by LOC to the observed one channel-by-channel and computing the $\chi^2$ for the minimisation. At each step, LOC is run with 50 iterations separately for the two isotopologues. The synthetic spectra are convolved to the angular resolutions listed in Table~\ref{tab:lines} and interpolated onto the spectral axes of the observations before computing the residuals. We first ran an initial simulation using 40 walkers for each free parameter and 500 steps. Then, we used the posterior distributions (in terms of medians and range) to select the initial set of parameters and their range for a second MCMC run (cf. Table~\ref{tab:MCMC}). The first one is then discarded as burn-in. Figure~\ref{fig:corner} shows the resulting corner plot, whilst Table~\ref{tab:MCMC} summarises the best parameter results. The code uses a logarithmic scale for the molecular abundance. \par
The synthetic spectra obtained with the median values of the free parameters are plotted in Fig.~\ref{fig:n2hp_n2dp}, and they generally agree with the observations. The intensities of the \nnhp lines are reproduced well. The model fails to reproduce the double-peak asymmetry seen in the $3-2$ line profile. This feature can be indicative of self-absorption combined with inward contraction motions. The fact that LOC cannot reproduce it 
could indicate that the physical model underestimates the contraction speed of the outskirts of the core, or that we undereste the \nnhp abundance in these regions, from where most of the $3-2$ flux likely arises. The observed linewidth of the \nnhp $5-4$ line (the broadest in the sample, cf. Table~\ref{tab:lines}) is underestimated by the model, a possible indication of increasing infall motions in the core's centre, as discussed in Sect.~\ref{sec:discussion}. The profiles of the \nndp lines are well reproduced. The peak intensities are slightly off, with the flux of the $3-2$ line underpredicted, whilst the $4-3$ and $6-5$ synthetic spectra are brighter than the observations. However, for the highest $J$ transition, the difference is comparable to the $rms$ of the observed data. 
\par
The best-fit solution prefers a factor $\sim 2$ denser physical structure than the model derived from the \textit{Herschel} continuum data and a higher (by $\sim 40\,$\%) gas temperature. The temperature at the core centre is hence closer to $11\, \rm K$, which is somewhat warmer than other typical pre-stellar sources. This could hint to the influence of the surrounding environment, as discussed further in Sect.~\ref{sec:discussion}. The main isotopologue abundance, $X \mathrm{(N_2H^+) } =1.00^{+0.07}_{-0.05}\times 10^{-10}$, is in line with estimates in similar sources (cf. \citealt{Hardegree-Ullman13} for CrA 151; \citealt{Crapsi05}, \citealt{Redaelli18} in other dense cores). The deuteration level is high: $R_\mathrm{D} =0.50^{+0.01}_{-0.01}$, which is expected due to the high densities and relatively low temperatures that favour CO depletion and deuteration processes. 

\begin{table}[!h]
    \renewcommand{\arraystretch}{1.4}
\centering
\caption{For each free parameter explored in the MCMC analysis, the second column reports the range explored (prior probability distribution). The third column states the median value obtained in the posterior distribution, together with the 0.16 and 0.84 percentiles as uncertainty range.}
\label{tab:MCMC}
\begin{tabular}{ccc}
\hline \hline
Parameter & Range & Value \\
\hline
$k_{n}$                &  $1.3-2.7$  &   $2.11^{+0.06}_{-0.06}$        \\
$X(\rm N_2H^+)/10^{-10}$&  $10^{-2}-10^{+2}$  &   $1.00^{+0.07}_{-0.05}$    \\
$k_{T \rm _K}$          &  $1.0-1.8$  &   $1.39^{+0.01}_{-0.01}$           \\
$R_\mathrm{D}$          &  $0.2-0.8$  &    $0.50^{+0.01}_{-0.01}$          \\
 \hline
\end{tabular}

\end{table}

\subsection{Other lines identified\label{sec:otherlines}}

\begin{figure*}
    \centering
    \includegraphics[width=.8\linewidth]{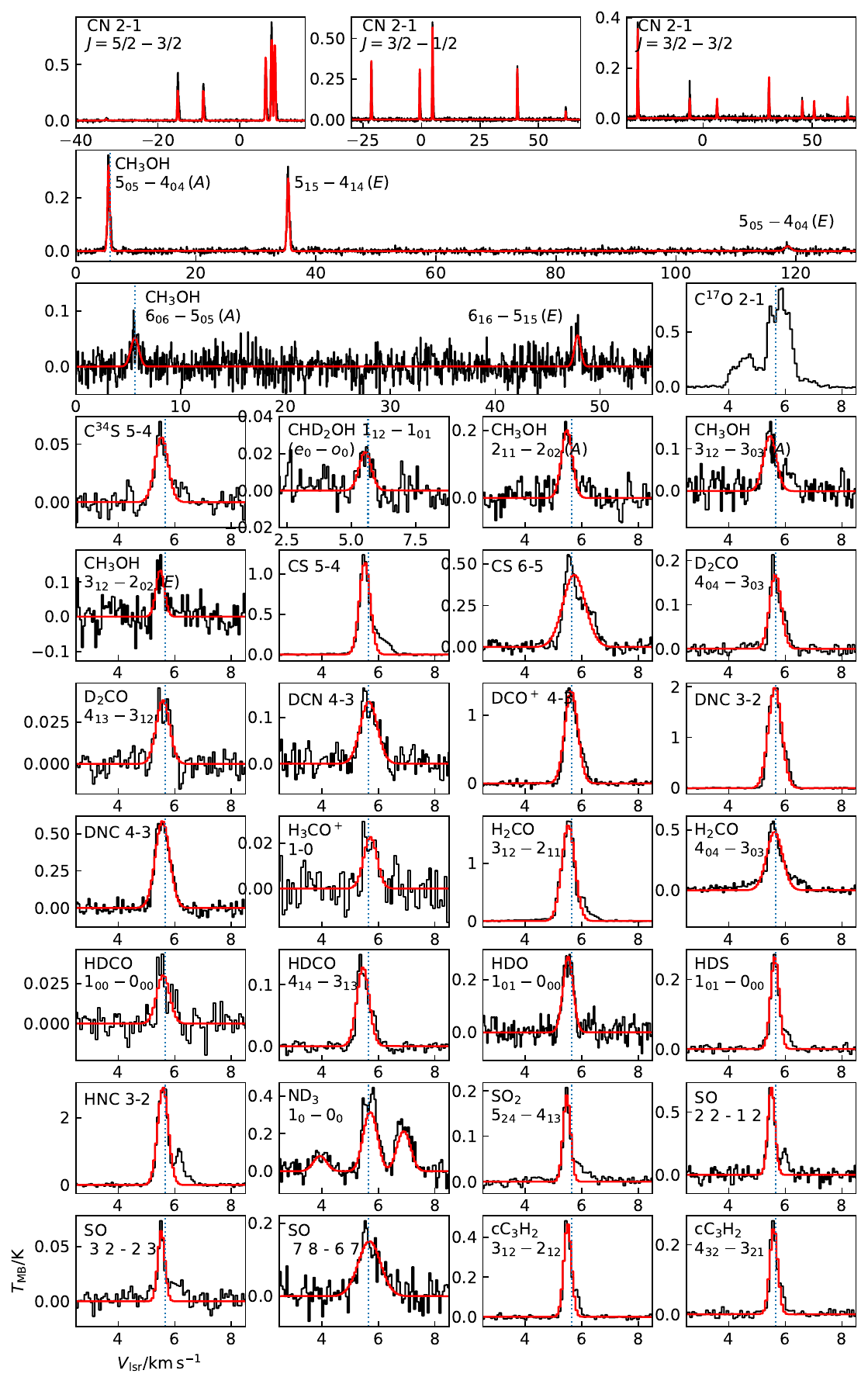}
    \caption{The black histograms show the observed spectra of the species identified in the frequency coverage of the \nnhp/\nndp data. The transitions are labelled in the top-left corner of each panel. The red histograms show the spectral fit performed with \textsc{class}, {which is a Gaussian fit for all transitions except for the $\rm ND_3$ and $\rm CN$ ones, for which we fit the hyperfine structure using the \textsc{hfs} routine}. The vertical dotted line in each panel shows the source $V_\mathrm{lsr} = 5.65\,$\kms, corresponding to the weighted average measured in the \nnhp and \nndp lines. }
    \label{fig:other lines}
\end{figure*}

The large bandwidth of the APEX backend ensured that the collected observations have a broad frequency coverage, covering the ranges $224.4-232.3\,$GHz, $240.6-248.5\,$GHz, $271.8-279.7\,$GHz, $287.3-295.9\,$GHz, $303.6-311.5\,$GHz, $462.2-466.2\,$GHz, and $474.6-478.6\,$GHz. We have searched the whole frequency coverage to identify other lines. The complete list of detected transitions above $\rm S/N = 3$ is given in Table~\ref{table:otherlines} and they are shown in Fig.~\ref{fig:other lines}. In total, we detected 35 transitions\footnote{We consider here a single value for the CN $J = 2-1$ line, for which however the whole fine structure pattern was observed, and for the dense hyperfine structure of $\rm ND_3 \; 1_{01}-0_{00}$ and $\rm C^{17}O \; 2-1$.} from 20 different species (including isotopologues). \par
All the nitrile transitions that fall in the covered frequency range have been detected: CN $2-1$, HNC $3-2$, DCN $4-3$, DNC $3-2$ and $4-3$. $\rm DCO^+$ 4-3 is also detected, with bright intensity ($T_\mathrm{peak} \sim 1.1\,$K). \par
Several transitions of methanol and the $J_{K_a K_c}$ = $1_{12}-1_{01} \, (e_0-o_0)$  of doubly deuterated methanol are detected, as well as two transitions for each deuterated isotopologue of formaldehyde. This confirms the high level of deuteration of the targeted core. We also detected triply deuterated ammonia $\rm ND_3$ and singly-deuterated water HDO. 
Concerning S-bearing species, we detected CS (and the rarer $\rm C^{34}S$), SO, SO$_2$, and $\rm HDS$. Finally, we detected cyclic $\rm C_3H_2$, protonated formaldehyde ($\rm H_3CO^+$), and C$^{17}$O.
\par
Using the \textsc{class} software, we performed Gaussian fits to the lines. In the case of CN and ND$_3$, we fitted only one hyperfine component; for C$^{17}$O 2-1, which presents three unresolved components, we did not perform the fit. The best-fit parameters are also summarised in Table~\ref{table:otherlines}.
\par
The detected transitions present different line shapes, which can be divided into two groups. The deuterated species are usually well-fitted by the single Gaussian component assumption. The main exceptions are the HDS line (which better fits among the sulphurated species, see the discussion below), and possibly the HDCO $4_{14}-3_{13}$ transition, where hints of a red shoulder can be seen. Species containing sulphur, instead, usually present a broad shoulder towards the higher velocities, up to $\sim 1\, $\kms above the core $V_\mathrm{lsr}$. This shoulder is also  identified in several other non-deuterated and non-sulphurated species, such as methanol, the main isotopologue of formaldehyde, CN and HCN. In some cases (e.g. HCN $3-2$, SO $2\,2 - 1\, 2$) this feature resembles a second velocity component located at $6.0-6.2 \,$\kms.

\begin{table*}{}
\caption{List of lines robustly detected (S/N>3) in the spectral ranges covered by the observations. For each line, the species, the type of transition, and the rest frequency (taken from the CDMS and JPL catalogues, \citealt{Muller05, Pickett98}) are given. The last three columns report parameters obtained from the Gaussian fit performed in \textsc{class}. Uncertainties on the last digit(-s) are written in brackets.}
\label{table:otherlines}
\centering
\begin{tabular}{cccccc}
\hline\hline \\[-2ex]
Species & Transition & Rest frequency\tablefootmark{a} &  W &  $V_\mathrm{lsr}$ & $FWHM$        \\
&       &MHz   & K km/s    & km/s & km/s                            \\[0.5ex]
\hline \\[-2ex]
$\rm CN$\tablefootmark{b} 			&   $ N \, J = 2 \, \frac{3}{2}  -  1 \, \frac{3}{2} $ 	&    226333.0904 	&   1.004(4) 		&   5.56691(3) 	&   0.391(4) \\
$\rm CH_3OH$\tablefootmark{c}		& $J_{K_a K_c}$ = $5_{05} -4_{04}\, (A)$ 	&241791.3520 			& 0.152(5)		& 5.491(7)		&0.44(2) \\
				& $J_{K_a K_c}$ = $5_{15}- 4_{14}\, (E)$ 	&241767.2340 	& 0.133(6) 	& 5.499(9)		&  0.46(3)\\
				&$J_{K_a K_c}$ = $5_{05}- 4_{04}\, (E)$	    & 241700.1590 		& 0.024(17) 	& 5.6(3) 		& 1.3(1.6) \\
$\rm CH_3OH$\tablefootmark{c}		&$J_{K_a K_c}$ = $6_{06} -5_{05}\, (A)$ 	& 290110.637 			&0.019(3) 		& 5.475(14) 		& 0.17(3) \\
				& $J_{K_a K_c}$ = $6_{16}- 5_{15}\, (E)$ 	& 290069.7470 		&0.042(8) 		& 5.56(5)		& 0.7(2) \\
$\rm CH_3OH$		& $J_{K_a K_c}$ = $2_{11} -2_{02}\, (A)$	& 304208.348		& 0.098(8)		&  5.501(17) 	&  0.46(5)\\
$\rm CH_3OH$		& $J_{K_a K_c}$ = $3_{12} -3_{03}\, (A)$ 	& 305473.491 			& 0.074(10)       & 5.49(3) 		& 0.53(11) \\
$\rm CH_3OH$		& $J_{K_a K_c}$ = $3_{12} -3_{202}\, (E)$ 	& 310192.994			& 0.050(8)  	& 5.52(3) 		& 0.35(7) \\
$\rm CHD_2OH$	&$J_{K_a K_c}$ = $1_{12}-1_{01} \, (e_0-o_0)$ 	& 241353.127			& 0.012(2) 	& 5.58(5) 		& 0.56(9) \\
$\rm H_2CO$ 		& $J_{K_a K_c}$ = $3_{12}- 2_{11}$ 	& 225697.7750	& 0.975(5)		& 5.4819(14)	& 0.552(3) \\
$\rm H_2CO$		& $J_{K_a K_c}$ = $4_{04}- 3_{03}$	& 290623.405		& 0.340(11) 	& 5.585(9) 	& 0.66(3) \\
$\rm HDCO$		    & $J_{K_a K_c}$ = $1_{11}- 0_{00}$ 	& 227668.453	 	& 0.017(3) 	& 5.55(5) 		& 0.54(14)\\
$\rm HDCO$ 		    & $J_{K_a K_c}$ = $4_{14}- 3_{13}$ 	& 246924.600		& 0.0709(18) 	& 5.496(6) 	& 0.519(16) \\
$\rm D_2CO$		    & $J_{K_a K_c}$ = $4_{04}- 3_{03}$ 		& 231410.234		& 0.087(2)		& 5.608(7)		& 0.485(17)	\\
$\rm D_2CO$		& $J_{K_a K_c}$ = $4_{13}- 3_{12}$  	& 245532.752			& 0.0228(16)	& 5.62(2)		& 0.56(4) \\
$\rm H_3CO^+$ 	&$J_{K_a K_c}$ = $1_{11}- 0_{00}$ 		& 226746.308			&0.0134(2) 	&5.68(4)		& 0.56(9) \\
$\rm c- C_3H_2$	& $J_{K_a K_c}$ = $4_{32} -3_{21}$		& 227169.1378		& 0.0879(19)		& 5.549(4)		& 0.331(10) \\
$\rm c- C_3H_2$	& $J_{K_a K_c}$ = $3_{21} -2_{12}$ 	& 244222.1505		& 0.159(2)		& 5.53(2) 		&0.321(5) \\
$\rm C^{17}O$\tablefootmark{d}		& $ J = 2-1$ 		& 224714.385 		&	-		&	-		&	-	\\
$\rm HNC $		    & $ J = 3-2$ 		& 271981.1420		&	1.39(4) & 5.548(5) & 0.450(15)\\
$\rm DCN	$		& $ J = 4-3$ 		& 289644.9170		& 0.101(6)		& 5.63(2) 		& 0.71(5)	\\
$\rm DNC$ 		    & $ J = 3-2$ 		& 228910.481		& 1.177(2) 	& 5.5936(5)	& 0.5584(13)\\
$\rm DNC$	      	& $ J = 4-3$ 		& 305206.219		& 0.372(9) 	& 5.593(7) 	& 0.596(16) \\
$\rm DCO^+$         & $ J = 4-3$        &288143.3858        &0.784(7) & 5.594(3) & 0.543(6) \\
$\rm CS$			& $ J = 5-4$		& 244935.5565 		& 0.546(9) 	& 5.559(3) 	& 0.449(10) \\
$\rm CS $			& $ J = 6-5$		& 293912.0865		& 0.430(8) 	& 5.696(10) 	& 0.93(2) \\
$\rm C^{34}S$		& $ J = 5-4$ 		& 241016.0892		& 0.033(2)		& 5.563(19)	& 0.55(5)	\\
$\rm SO$	 		&  $N \, J = 3\, 2 - 2\, 3$		& 246404.5881 		& 0.0203(18) 	& 5.559(10) 	& 0.29(4) \\
$\rm SO$			& $ N\, J = 7\,8 - 6\, 7$		& 304077.844		& 0.150(11)	& 5.71(3)		& 0.93(9) \\
$\rm SO$		& $ N \,J = 2\, 2 - 1\, 2$		& 309502.444			& 0.219(10) 	& 5.537(6) 	& 0.296(17) \\
$\rm SO_2$ 		& $J_{K_a K_c}$ = $5_{24}- 4_{13}$  	& 241615.7967		& 0.062(2)		& 5.505(5)		& 0.306(14) \\
$\rm HDS$		& $J_{K_a K_c}$ = $1_{01}- 0_{00}$  	& 244555.580		& 0.099(2)		& 5.666(3) 	&0.352(9) \\
  ND$_3$\tablefootmark{e} 			&   $J_{K} =  1_ {0} - 0_{0}$  	&  309909.490		&  0.78(10) 	&  5.748(9) 	&   0.53(2) \\
HDO 			& $J_{K_a K_c}$ = $1_{01}- 0_{00}$  	& 464924.520	 	& 0.129(6) 	& 5.493(10) 	& 0.42(2)  \\
\hline
\end{tabular}
\tablefoot{ \tablefoottext{a}{Rest frequencies are derived from the laboratory work in \cite{bocquet99, bechtel06, clark76, helminger69, messer84, cazzoli02, klapper03, dixon77, caselli05, helminger71, bruenken06, saykally76, melosso21, zakharenko15, mueller17, xu08, bogey86}.}\\
\tablefoottext{b} { {  We report the \textsc{class hfs} fit results for the $N \, J = 2 \, \frac{3}{2}  -  1 \, \frac{3}{2} $ transition. The first best-fit parameter is the $\Sigma T_{\rm ant} \times \tau$ quantity.}} \\
\tablefoottext{c} {The transitions are shown in a single panel in Fig.~\ref{fig:other lines}.}\\
\tablefoottext{d} {The hyperfine structure is barely resolved and the Gaussian fit is not performed.}\\
\tablefoottext{e} {   The entire $ J_{K} = 1_0 - 0_0$ is fitted via the \textsc{hfs} fit of \textsc{class}. The first best-fit parameter is the $\Sigma T_{\rm ant} \times \tau$ quantity.}}
\end{table*}

\section{Discussion \label{sec:discussion}}
Corona Australis 151 is a core well known for its high central density. \cite{bresnahan18} estimated an average density $\langle n \rangle = 10^7 \, \rm cm^{-3}$ in the central 1000$\,$au. This is confirmed by the detection of the high $J$ transitions of \nnhp and \nndp that we analyzed in this work, {and by the fact that with our non-LTE models, we cannot reproduce both the low- and high-$J$ transitions of \nnhp and \nndp with volume densities lower than few 10$^7$ cm$^{-3}$, as shown in Appendix~\ref{app:tests}}. In particular, the \nnhp $5-4$ and \nndp $6-5$ lines, observed with a linear resolution of $\approx 2000 \, $au, have critical densities of $4-5\times 10^6 \, \rm cm^{-3}$. 
\par
{Our core model derived from {\sl Herschel} maps at 160, 250, 350, and $500\,\mu$m is in agreement with} this picture, with densities higher than $10^6  \, \rm cm^{-3}$ within the central $5\times 10^{-3} \, \rm pc = 1000 \, au$, reaching a maximum of $3.2 \times 10^7  \, \rm cm^{-3}$ at $\sim 150 \,$au. The average density within the central 1000$\,$au is $\sim 6 \times 10^6 \, \rm cm^{-3}$.
At the same time, the derived dust temperatures are low ($T_\mathrm{dust} < 12 \, \rm $K), which is consistent with pre-stellar gas in a quiescent environment. However, based on the available continuum data, we estimate a bolometric temperature for the core of $T_\mathrm{bol}= 18.7\, \rm K$, excluding significant internal heating from a protostellar source (Caselli et al., in prep). The MCMC radiative transfer modelling of the lines is consistent with the high central density of the core ($n \approx 6 \times 10^7 \, \rm cm^{-3}$). 
\par
The comparison between the observational properties of dense cores and numerical simulations suggests that the dynamical evolution of an isolated prestellar core can be approximated by a sequence of hydrostatic cores in unstable equilibria, the so called thermally supercritical Bonnor-Ebert spheres \citep[BES,][]{Foster93, Keto10, Keto15}. The density profile of a supercritical BES approaches the power-law $n \propto r^{-2}$ at large distances from the core centre, extending inwards to a point called the "flat radius", which marks the radius where the sound crossing time equal the free-fall time. The flattened density region shrinks with time while the central density increases{, as also shown in ALMA observations of pre-stellar cores \citep{tokuda20}}. In the marginally stable model, called the quasi-equilibrium (QE) BES model by \cite{Keto15}, the infall velocity reaches its maximum just outside the flat radius, and goes to zero at the centre of the core and in the outer envelope. 
{This feature clearly distinguishes the QE BES model from other collapse models, which often predict similar $\sim r^{-2}$ density profiles, but have very different infall velocity profiles. The QE BES model has been successfully used to interpret molecular line observations towards the prestellar core L1544 \citep{Keto15,Redaelli19, Redaelli21, ChaconTanarro19,Rawlings24}. \cite{Keto15} also discussed a non-equilibrium (NE) BES model, which was constructed by increasing the density everywhere in an unstable BES, causing the sphere collapse on all scales. The infall velocity profile adopted here in the modelling of the observed spectra (Sect.~\ref{sec:LOC}) represents a crude approximation of the velocity distribution predicted by the NE BES model of \cite[][see their Fig. 3]{Keto15}.} The fact that the \nnhp $5-4$ and \nndp $6-5$ lines are detected in CrA 151 but not in L1544 (Caselli et al., in prep.) {strongly suggests} that CrA 151 has reached a higher central density than L1544. {While it is possible to reproduce the high-$J$ transitions of both N$_2$H$^+$ and N$_2$D$^+$ with lower volume densities ($3-6 \times  10^6 \rm \, cm^{-3}$) and higher \nnhp and \nndp abundances, it is not possible to reproduce $both$ low- and high-$J$ transitions simultaneously without volume densities as high as 10$^7$ cm$^{-3}$ (see also Appendix~\ref{app:tests}).}
The density profile of CrA 151 shown in Fig.~\ref{fig:cra151_model}, agrees with the $r^{-2}$ power law, except for the depression around $r=0.01 \,$pc, which is accompanied by a bump in the dust temperature profile. These features are invoked by corresponding bumps in the intensity profiles at 160 and 250$\, \mu$m. The irregularities may be understood in terms of deviations from the idealised models described above, where particles are always accelerated towards the centre, and no shocks occur when the infall speed exceeds the sound speed \citep{Foster93}.\par
{CrA 151, then, probably} represents a more advanced stage of prestellar core evolution than L1544.
We have {three} pieces of evidence in support of this hypothesis. Firstly,{there is the detection of compact $70\,\mu$m emission at the centre of the core, with a flux density of $0.15\pm0.03$\,Jy \citep{bresnahan18}}. Secondly, the increasing linewidth of \nnhp and \nndp lines with increasing frequency, which points toward faster motions in the central, denser regions of the core. Thirdly, several lines present high-velocity wings (up to $\sim 1\,$\kms above the source $V \rm _{lsr}$) (see Sect.~\ref{sec:otherlines}). {With the present data from a single pointing, we cannot rule out the possibility that the high-velocity wings are caused by external influence, for example by an outflow from the nearby protostar lying approximately 0.4$\,$pc to the North-West of the core \citep[][see also Panel a of Fig.~\ref{fig:HerschelData}]{Esplin22}. To our knowledge, no search for protostellar outflows has been conducted in this region, and we cannot exclude that this source is launching an outflow that is affecting the properties of CrA 151. Another possibility, however, is the presence of supersonic infall motions. In these regards, we tested a velocity profile that includes them, and the results are compatible with the observed lines of \nnhp and \nndp. However, observations at a higher resolution, as well as a dedicated model, are needed to confirm the presence of a supersonic infall. Although our MCMC analysis does not suggest the need of higher central temperatures to reproduce the observed lines of \nnhp and \nndp, and the very large deuterium fractionation measured also supports the pre-stellar core option, it is possible that all the three pieces of evidence listed above are manifestations of a very young protostar.}

\par
{A greybody fit to the flux densities from 70 to $850\,\mu$m
(combining the data from the HGBS archive with the $850\,\mu$m
aperture flux from SCUBA, $1.6\pm0.1$\,Jy) gives a bolometric
temperature of 18.7\,K and bolometric luminosity of $0.1\,L_\odot$ for
the core. The latter value is an upper limit for the luminosity of the
possible central source, which, therefore, should be classified as a
VeLLO.  The $70\,\mu$m map was excluded in the derivation of the core
model, because the compact, low-intensity source is smoothed out when
convolved with the {\sl Herschel} $500\,\mu$m beam. We tested how the
model should be modified in order to reproduce the observed $70\,\mu$m
map. Adopting the density structure of our model but assuming a
uniform temperature throughout the core, the best fit to the observed
map was obtained with $ T=14\rm \,K$. This suggests that the temperature is
indeed elevated in the very centre, although lower resolution data
give there 8\,K (3D model) or 10\,K ($T_{\rm C}$ from {\sl
Herschel}). In view of severe obscuration from the interstellar
radiation field, a central temperature of 14\,K would imply internal
heating, either by shocks in the accreting material or outflows. In
case the core contains a FHSC, the central density should exceed
$10^{-12}\,{\rm g\, cm^{-3}}$ corresponding to $n({\rm
H_2})\sim10^{11}\,{\rm cm^3}$ (e.g. \citealt{young19}). The spectral wings observed for the sulphurated molecules may trace a low-velocity outflow, which is
expected from a FHSC \citep{Machida08,Fujishiro20}.}
\par
The high central densities cause CO depletion, as investigated by \cite{Hardegree-Ullman13}. In turn, the depletion of CO molecules onto dust grains, combined with the low temperature, increases the efficiency of deuteration processes, as the production of $\rm H_2D^+$ is favoured. The high level of deuteration of CrA 151 is confirmed by the analysis of the diazenylium lines. The radiative transfer analysis implies a deuteration level of 50\%. For comparison, the \nndp/\nnhp ratio in L1544 is $26$\% \citep{Redaelli19}; \cite{Friesen13} found $R_\mathrm{D} \lesssim 0.20$ in a sample of starless cores in Perseus; in the L1688 region in Ophiuchus, \cite{Punanova18} reported deuteration fraction in the range $2-40$\%. In a twin work dedicated to another of the densest cores in the sample (Oph 464, or IRAS16293E), \cite{spezzano24} found values as high as 44\%. The high deuteration level of CrA 151 is corroborated by the number of deuterated species detected (see Sect.~\ref{sec:otherlines}), including doubly ($\rm CHD_2OH, D_2CO$) and triply ($\rm ND_3$) deuterated isotopologues. We also report the first detection of deuterated water in a pre-stellar environment.
\par
\par
The modelling we performed bears several limitations. The first one is the usage of a spherically symmetric model. The dust continuum maps shown in Fig.~\ref{fig:HerschelData} clearly deviate from spherical symmetric around the core's centre. Nevertheless, this approximation seems reasonable within the central $\sim 2'$, which is comparable to the simulated core radius ($0.14\, \rm pc$ or $190''$). Furthermore, the development of a fully three-dimensional core structure goes beyond the scope of the present paper, which intends to investigate the physical conditions at the core's centre. The MCMC analysis could also be expanded by introducing more free parameters such as, for instance, the velocity profile or the turbulence contribution. However, increasing the number of free parameters whilst maintaining the same number of observational constraints (the five \nnhp and \nndp lines) does not guarantee a better convergence of the method, as discussed also by \cite{Jensen24}. Therefore, we preferred to focus on the parameters that are more relevant to our analysis: density, temperature, and the abundance of the two species. 
\par
It is worth commenting on the choice of constant abundance for \nnhp and \nndp. As discussed in Sect.~\ref{sec:LOC}, these profiles reproduce reasonably well the line profiles and the intensity ratio of the distinct $J$ transitions, even though there is a tendency to overestimate the fluxes of the high-frequency transitions of \nndp (by 200$\,$mK for \nndp $4-3$, and by 50$\, \rm mK$ for \nndp $6-5$; the $rms$ levels of the two lines are 9$\,$mK and 29$\,$mK, respectively). Since the high-$J$ transitions have higher critical density and thus are emitted closer to the core's centre, the modelling could benefit from some level of depletion in the central parts of the core. However, the situation is different from the case of L1544, where if the \nndp abundance profile is kept constant in the central $\approx 4000\, $au, the higher-$J$ transitions are overestimated by a factor of 2 or more compared to the low-$J$ one \citep{Redaelli19}\footnote{In that paper the authors analysed \nndp $1-0$, $2-1$, and $3-2$. The $4-3$ and $6-5$ transitions are not detected in L1544 (Caselli et al., in prep.).}. CrA 151 resembles instead the situation of IRAS16293E analyzed by \cite{spezzano24}, where the authors concluded that models with either no depletion, or with a small central depletion zone of $\lesssim 1000\, $au can reproduce the observations. \par
These findings are in disagreement with chemical models, which in general predict some level of N$_2$ (and thus \nnhp and \nndp) depletion at high densities (cf. \citealt{Sipila19}) and low temperatures. The limited angular resolution of our observations might play a role. In L1544, the depletion of \am is detected only with ALMA high-resolution (400$\,$au) observations \citep{Caselli22}. Furthermore, the core temperatures derived with the MCMC analysis in CrA 151 ($12-16\,\rm K$) are higher than those detected in other quiescent cores, where $T_\mathrm{gas} \lesssim 10 \, $K \citep{Crapsi07, Lin23}.

\section{Summary and Conclusions \label{sec:conclusions}}
We reported the detection of the \nnhp $3-2$ and $5-4$ and the \nndp $3-2$, $4-3$, and $6-5$ lines towards the dense core CrA 151 using the APEX single-dish telescope. This core has been selected for further investigation and modelling from the sample of dense starless sources collected from the \textit{Herschel} HGBS catalogue (Caselli et al., in prep.). Analysing the \textit{Herschel} continuum maps, we built a spherical model of the source which is characterised by high central densities ($n\gtrsim 10^7\,\rm cm^{-3}$) and overall low dust temperatures ($8-12\, \rm K$). \par

Starting from a 1D model, we modelled the spectra using non-LTE radiative transfer  coupled with MCMC analysis, using the abundance of \nnhp and its deuterium fraction as free parameters. We also let the density and temperature profiles vary by constant factors. This modelling predicts that even higher densities (a factor of $\sim 2$), and warmer temperatures (factor of $\sim 1.4$) than those derived from the \textit{Herschel} fluxes are needed to reproduce the observations. Constant abundance profiles of $X(\rm N_2H^+) \sim 10^{-10}$ and $X(\rm N_2D^+) \sim 5\times 10^{-11}$ are in reasonable agreement with the APEX spectra, even though we cannot exclude some level of depletion limited to the central $\approx 1000\, $au. The derived deuterium fraction is $50 \%$. This deuteration, enhanced by the high densities and relatively low gas temperature, is among the highest levels seen in similar sources. 
\par
In the large frequency coverage of the APEX data we detected 20 other species, including several deuterated isotopologues. The detection of $\rm CHD_2OH$, $\rm D_2CO$, and $\rm ND_3$ highlights the generally high levels of deuteration of the source. These transitions are well-fitted by single Gaussian components with narrow linewidths ($FWHM \lesssim 0.5\,$\kms). This evidence supports the scenario of a dense, cold and quiescent core. However, the other detected species, in particular sulphur-bearing ones, often show line profiles with extended wings at high velocities ($V_\mathrm{lsr} > 6\,$\kms). There are {three} possibilities to interpret these profiles: \textit{i)} possible influence of the surrounding environment, where at least one protostar has been identified at a projected distance of $0.4\, \rm pc$; {\textit{ii)} presence of a very young embedded protostar or a FHSC}; \textit{iii)} the presence of supersonic infall motions in the central parts of the core. The latter scenario is supported by the fact that, for diazenylium, we observe an increase of intrinsic linewidth as the critical density and angular resolution of the transitions increase, by the high level of deuterium fractionation measured for \nnhp, and by the physical modelling of the source, which points towards an evolved dynamical pre-stellar core. 
\par
{The spectral features mentioned above and the detection of faint
70$\, \mu$m emission suggest the presence of an internal heating source
in CrA\,151. With an upper limit of $0.1\,L_\odot$, the hypothetical
central source would fall in the category of Very Low Luminosity
Objects (VeLLOs), and is thereby also a candidate First Hydrostatic
Core (FHSC). CrA\,151 may represent the very short and elusive phase
when a prestellar core is transforming itself into a protostellar
core. As an isolated, nearby object, CrA\,151 is an ideal target for
future high-resolution (tens to hundreds au) continuum and spectral
line observations to investigate the innermost structure of a dense
core, the kinematics of the infalling gas, and, if the core already
contains a protostar or a FHSC, the earliest stages of a molecular
outflow.}
\begin{acknowledgements}
   The data was collected under the Atacama Pathfinder EXperiment (APEX) 
Project, led by the Max Planck Institute for Radio Astronomy at the ESO La 
Silla Paranal Observatory. This research has made use of data from the {\em Herschel} Gould Belt survey (HGBS) project (\url{http://gouldbelt-herschel.cea.fr}). The HGBS is a Herschel Key Programme jointly carried out by SPIRE Specialist Astronomy Group 3 (SAG 3), scientists of several institutes in the PACS Consortium (CEA Saclay, INAF-IFSI Rome and INAF-Arcetri, KU Leuven, MPIA Heidelberg), and scientists of the Herschel Science Center (HSC). This research has made use of spectroscopic and collisional data from the EMAA database (https://emaa.osug.fr and https://dx.doi.org/10.17178/EMAA). EMAA is supported by the Observatoire des Sciences de l’Univers de Grenoble (OSUG).

\end{acknowledgements}

%
%

\begin{appendix}

    \section{\textsc{class hfs} fit\label{app:CLASSFit}}
\begin{figure*}[!t]
    \centering
    \includegraphics[width=0.8\textwidth]{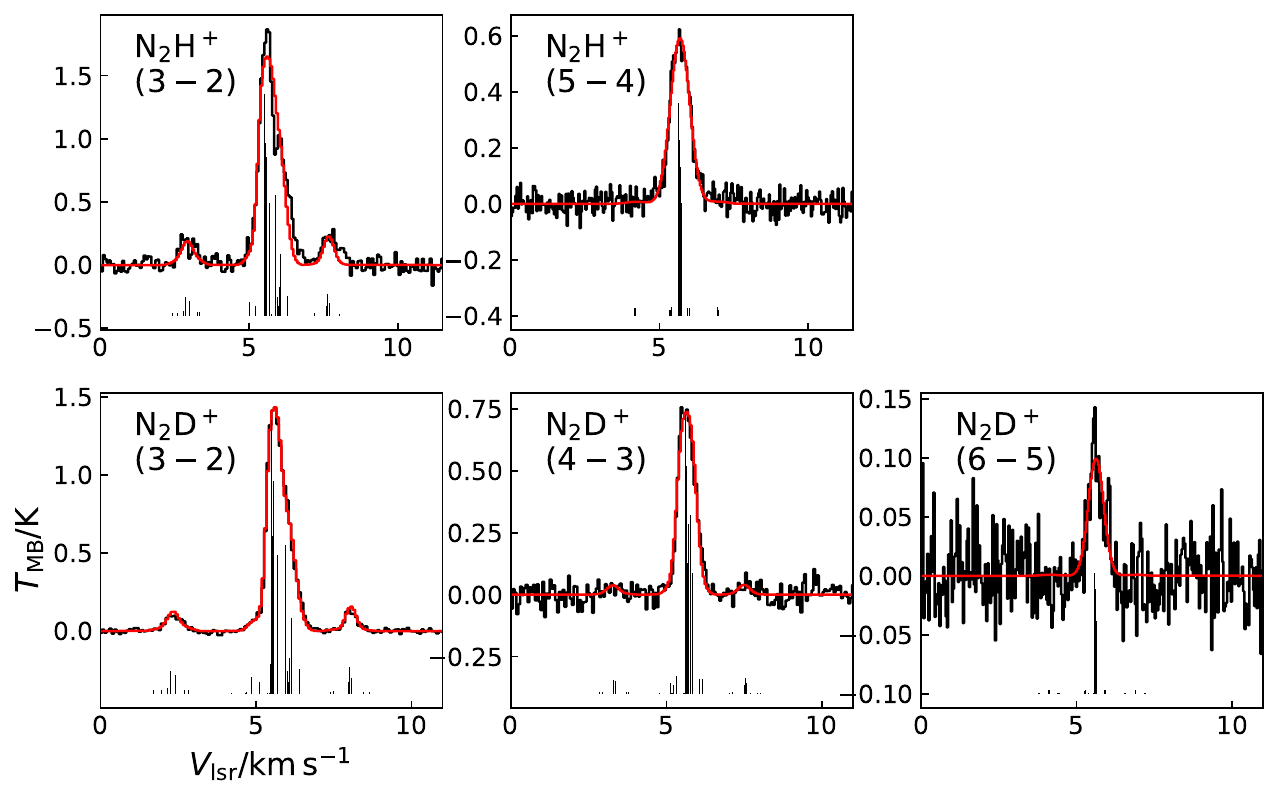}
\caption{The black histograms show the \nnhp and \nndp spectra, as in Fig.~\ref{fig:n2hp_n2dp}. The red ones show the best-fit solution found by \textsc{class}, assuming the same excitation temperature for all the hyperfine components. The vertical, black lines show the position of the hyperfine components, their length proportional to the relative intensities. \label{fig:CLASSFit}}
\end{figure*}
In Fig.~\ref{fig:CLASSFit} we show the best-fit solutions found with the \textsc{class hfs} fitting routine, as described in Sect.~\ref{sec:results}.

\section{\text{Herschel} maps \label{app:Herschel}}
\begin{figure*}[!h]
    \centering
    \includegraphics[width=\textwidth]{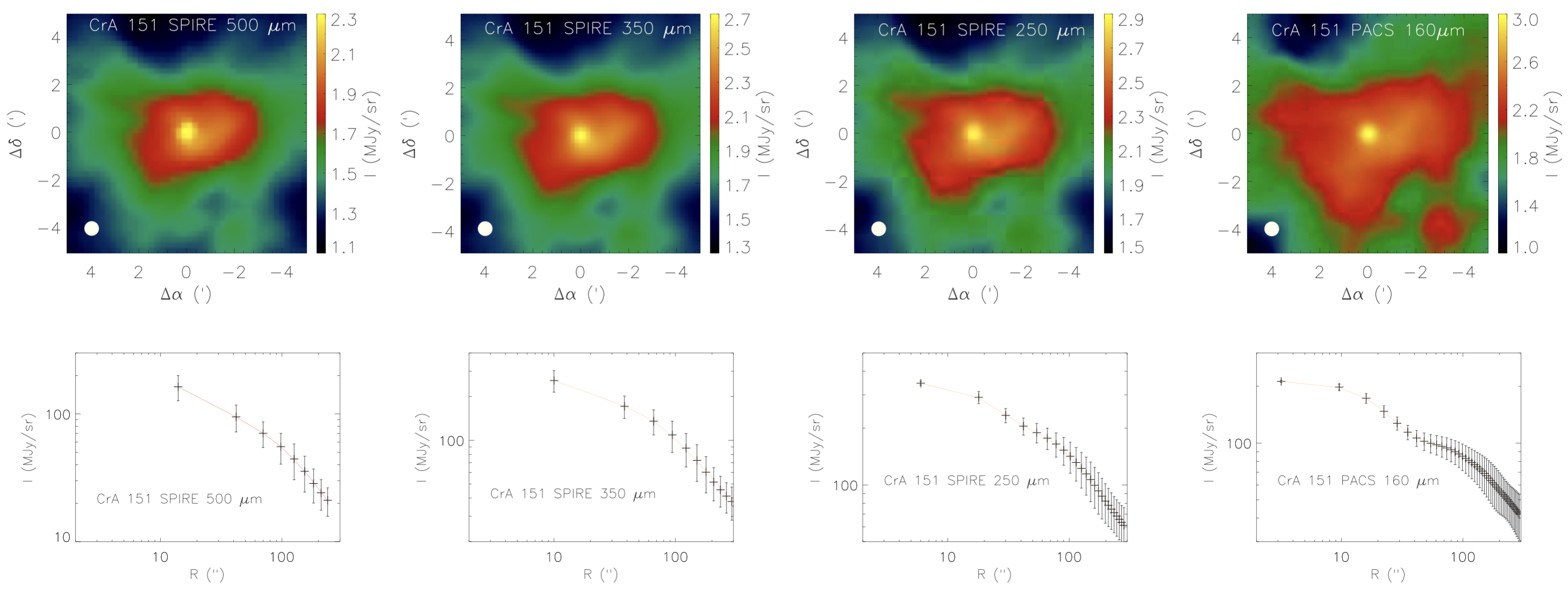}
\caption{  Top row: \textit{Herschel} SPIRE and PACS maps; Bottom row: Corresponding circularly averaged intensity profiles. One can note the broader emission at large radii in the 160 and 250$\, \mu$m maps that require two Plummer profiles for an optimal fit. The 160, 250, and 350$\,\mu$m maps have been smoothed to the resolution of the $500\,\rm \mu m$ one. \label{fig:herschel_maps}}
\end{figure*}

  Figure~\ref{fig:herschel_maps} shows the \text{Herschel} data of CrA 151 from the SPIRE and PACS instruments in the top row. The corresponding bottom panels show the intensity profiles, circularly averaged around the core's centre.

 \section{LOC tests\label{app:tests}}
\begin{figure*}[!h]
    \centering
    \includegraphics[width=0.8\textwidth]{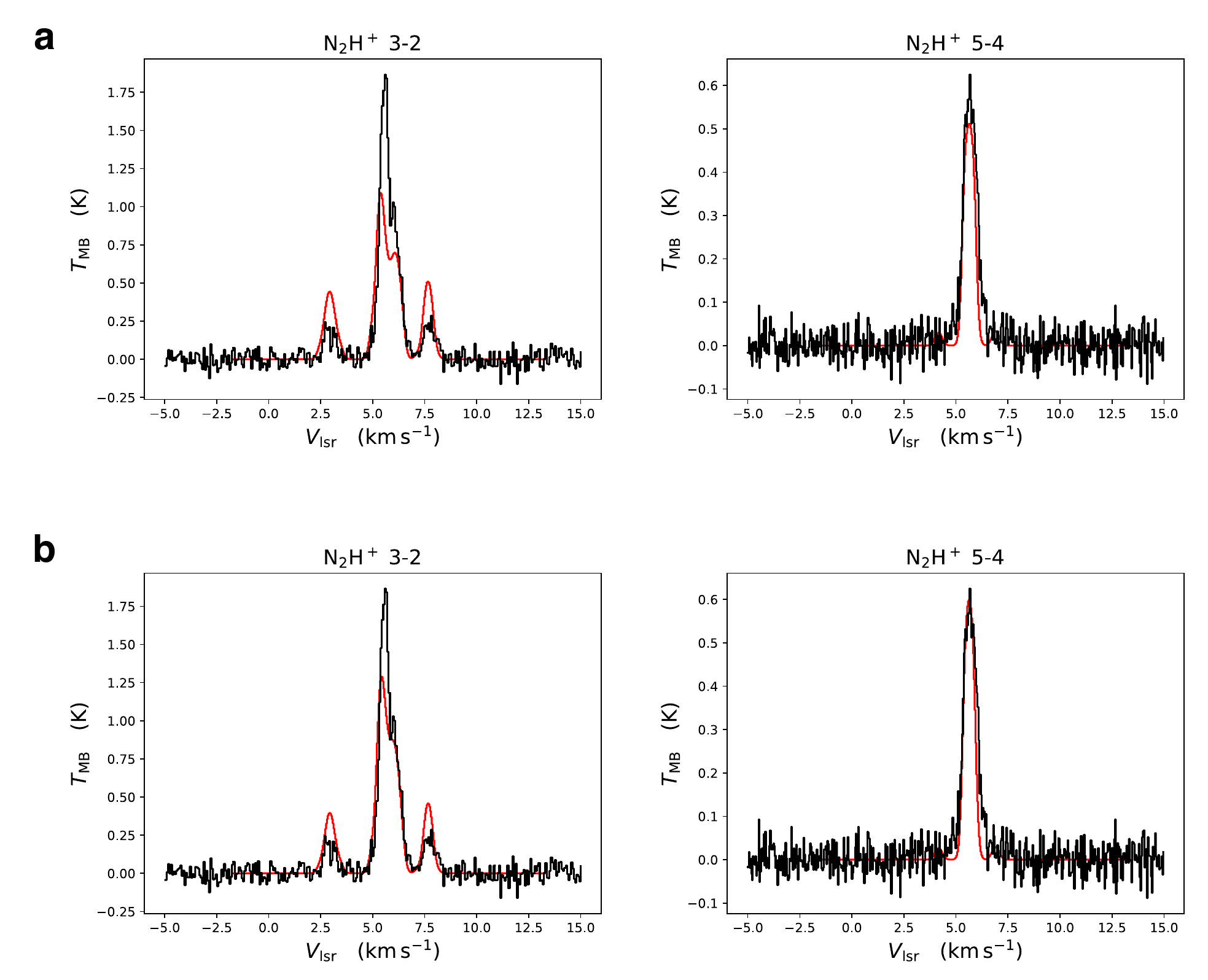}
\caption{   LOC tests with lower volume density. Panel (a): we scaled down the volume density profile used for the results shown in Fig.~\ref{fig:n2hp_n2dp} by a factor of 10 and scaled up the abundance profile of N$_2$H$^+$ by a factor of 45 in the effort of reproducing the observed lines. Panel (b): we reduced the volume density profile used in Fig.~\ref{fig:n2hp_n2dp} by a factor of 5 and scaled up the abundance profile of N$_2$H$^+$ by a factor of 15. In both cases, we show that it is not possible to reproduce both transitions with a lower volume density by only increasing the abundance of N$_2$H$^+$. \label{fig:LOC_test}}
\end{figure*}
 
Figure~\ref{fig:LOC_test} shows the output of LOC for N$_2$H$^+$ using the input profiles of the best model shown in the paper (cf. Fig.~\ref{fig:n2hp_n2dp}). Here, we have reduced the volume density by a factor of 10 (upper panel) and 5 (lower panel), and increased the N$_2$H$^+$ abundance accordingly to find a good match with the observations. The results show that if the volume density is not high enough, the column densities necessary to reproduce the high-$J$ transition are too high for the low-$J$ transitions.

\end{appendix}
\end{document}